\documentclass[onecolumn,noshowpacs,noshowkeys,pra,longbibliography,nofootinbib,notitlepage]{revtex4-1}%
\usepackage[caption=false]{subfig}
\usepackage{algorithm}
\usepackage{color}
\usepackage{algorithmic}
\usepackage{amsmath}
\usepackage{amssymb}
\usepackage{bm}
\usepackage{array}
\usepackage{enumerate}
\usepackage{url}
\usepackage{graphicx}
\usepackage{bibentry}

\providecommand{\U}[1]{\protect\rule{.1in}{.1in}}

\DeclareMathOperator{\erf}{Erf}

\newcommand\myatop[2]{\genfrac{}{}{0pt}{}{#1}{#2}} 

\begin{document}
\title{Quantum Key Distribution Using Multiple Gaussian Focused Beams}
\thanks{This document does not contain technology or technical data controlled under either the U.S.~International Traffic in Arms Regulations or the U.S.~Export Administration Regulations. 
BAB acknowledges the support of the SECANT program funded by the Sandia National Laboratories under PO\# 1628276. 
NC and JHS acknowledge support from the Air Force Office of Scientific Research (AFOSR) grant number FA9550-14-1-0052.
BAB and SG acknowledge the support from the SeaKey program, through the US Office of Naval Research (ONR) contract number N00014-14-C-0002.}
\author{Boulat A. Bash$^{1}$ and Nivedita Chandrasekaran$^{2}$ and Jeffrey H. Shapiro$^{2}$ and Saikat Guha$^{1}$}
\affiliation{$^{1}$Quantum Information Processing Group, Raytheon BBN Technologies, Cambridge, MA 02138, USA\\
$^{2}$Research Laboratory of Electronics, Massachusetts Institute of Technology, 77 Massachusetts Avenue, Cambridge, MA 02139, USA
}

\begin{abstract}
The secret key rate attained by a free-space QKD system in the {\em near-field} propagation regime (relevant for $1$-$10$ km range using $\approx 7$ cm radii transmit and receive apertures and $1.55~\mu$m transmission center wavelength) can benefit from the use of multiple spatial modes. A suite of theoretical research in recent years has suggested the use of orbital-angular-momentum (OAM) bearing spatial modes of light to obtain this improvement in rate. We show that most of the aforesaid rate improvement in the near field afforded by spatial-mode multiplexing can be realized by a simple-to-build overlapping Gaussian beam array (OGBA) and a pixelated detector array. With the current state-of-the-art in OAM-mode-sorting efficiencies, the key-rate performance of our OGBA architecture could come very close to, if not exceed, that of a system employing OAM modes, but at a fraction of the cost.
\end{abstract}
\maketitle

\section{Introduction}

The extremely low key rates afforded by quantum key distribution (QKD) compared to computational cryptographic schemes pose a significant challenge to the wide spread adoption of QKD. The main reason for the poor rate performance is that the QKD {\em capacity} of a single-mode lossy bosonic channel, i.e., the maximum key rate attainable using any direct-transmission QKD protocol, is proportional to the end-to-end transmissivity of the channel $\eta$ in the high-loss regime. Therefore, to increase the key rate one must increase the number of modes used by the system.  This can be done by increasing the optical bandwidth $\nu$ in modes/s that can be used by the QKD protocol as well as employing multiple spatial modes.  Here we investigate the latter.

Formally, the QKD capacity of a single-mode bosonic channel where we employ both polarizations of light is $2\nu \log_2\left(\frac{1}{1-\eta}\right)\approx 2.88\,\nu\eta$ bits/s when $\eta \ll 1$ \cite{pirandola15QKDcap}..  Since $\eta \propto \frac{e^{-\alpha L}}{L^2}$, this corresponds to an exponential decay of key rate with distance $L$ in fiber and free-space propagation in non-turbulent atmosphere.  While the extinction coefficient $\alpha$ may be modest for the atmospheric propagation in clear weather at a well-chosen wavelength, the inverse-square decay of rate with distance is unavoidable in the {\em far-field} regime even in vacuum (where $\alpha=0$). This is because free-space optical channel is characterized by the Fresnel number product $D_{\rm f} \equiv A_{\rm t}A_{\rm r}/(\lambda L)^2$, where $A_{\rm t}$ and $A_{\rm r}$ are the respective areas of the transmitter and receiver apertures, and $\lambda$ is the transmission center wavelength. In the far-field regime $D_{\rm f} \ll 1$ and only one transmitter-pupil spatial mode couples significant power into the receiver pupil over an $L$-meter line-of-sight channel with input-output power transmissivity $\eta_0 \approx D_{\rm f} \propto 1/L^2$ \cite{Sha05}. Thus, employing multiple orthogonal spatial modes in the far-field regime cannot yield a appreciable improvement in the achievable QKD rate. 

Therefore, our interest in this paper is in the {\em near-field} propagation regime ($D_{\rm f} \gg 1$), which is relevant to metropolitan area QKD, as well as line-of-sight over-the-surface maritime applications of QKD. In this near-field regime, approximately $D_{\rm f}$ mutually-orthogonal spatial modes have near-perfect power transmissivity ($\eta \approx 1$) \cite{Sha05}.  Thus, multiplexing over multiple orthogonal spatial modes could substantially improve the total QKD rate with the gain in rate over using a single spatial mode (such as a focused Gaussian beam) being approximately proportional to $D_{\rm f}$, and hence more pronounced at shorter range $L$ (where $D_{\rm f}$ is high).

Laguerre-Gauss (LG) functions in the two-dimensional transverse coordinates form an infinite set of mutually-orthogonal spatial modes, which happen to carry orbital angular momentum (OAM). There have been several suggestions in recent years to employ LG modes for QKD, both based on laser-light and single-photon encodings~\cite{berkhout10oam,mirhosseini15twistedlightqkd,horiuchi15twistedbeam,
  vallone14twistedlightqkd,krenn14vienna,malik12oamturb,djordjevic13oamqkd,
  junlin10sixstate}, and the purported rate improvement has been attributed to the OAM degree of freedom of the photon. While multiplexing over orthogonal spatial modes could undoubtedly improve QKD rate in the near-field propagation regime as explained above: 
\begin{enumerate}[{(1)}]
\item Can other orthogonal spatial mode sets that do {\em not} carry OAM be as effective as LG modes in achieving the spatial-multiplexing rate improvement in the near field?
\item Does one truly need orthogonal modes to obtain this spatial-multiplexing gain or are there simpler-to-generate mode profiles that might suffice? 
\end{enumerate}
  
Question (1) was answered affirmatively for classical \cite{Sha05, chandrasekaran14pieturb1} and quantum-secure private communication (without two-way classical communication as is done in QKD)~\cite{chandrasekaran14pieturb2} over the near-field vacuum propagation and turbulent atmospheric optical channels: Hermite-Gauss (HG) modes are unitarily equivalent to the LG modes and have identical power-transfer eigenvalues $\left\{\eta_m\right\}$, $1 \le m < \infty$. Since the respective communication capacity of mode $m$ is a function of $\eta_m$ and the transmit power on mode $m$, HG modes, which do {\em not} carry OAM, can in principle achieve the same rate as LG modes, notwithstanding that the hardware complexity and efficiency of generation and separation of orthogonal LG and HG modes could be quite different. 

Our goal is to address questions (1) and (2) above for QKD. The answer to (1) is trivially affirmative, at least for the case of vacuum propagation (no atmospheric turbulence or extinction), based on an argument similar to the one used in Refs.~\cite{Sha05, chandrasekaran14pieturb1, chandrasekaran14pieturb2}.  We show  potential gain of between $1$ to $2$ orders of magnitude in the key rate by using multiple spatial modes over a $1$ km link, assuming $\approx 7$ cm radii transmitter and receiver apertures, and $\lambda = 1.55 \mu$m laser-light transmission.

The bulk of our analysis addresses question (2) for the optical vacuum propagation channel, which we answer negatively. We show that most of the spatial-multiplexing gain afforded by mutually-orthogonal modes (either HG or LG) in the near field can be obtained using a focused overlapping Gaussian beam array (OGBA) with optimized beam geometry in which beams are individually amplitude and/or phase modulated to realize the QKD protocol. 
These Gaussian focused beams (FBs) are {\em not} mutually-orthogonal spatial modes, and therefore the power that leaks into FB $m$ from the neighboring FBs has the same effect on the key rate $R_m(L)$ of that FB as do excess noise sources like detector dark current or electrical Johnson noise. Non-zero excess noise causes the rate-distance function $R_m(L)$ to fall to zero at a minimum transmissivity threshold $\eta_{\rm min}$, or, equivalently, at a maximum range threshold $L_{\rm max}$ such that $R_m(L) = 0$ for $L>L_{\rm max}$. Thus, while packing the FBs closer increases the spatial-multiplexing gain, it also increases the excess noise on each FB channel, resulting in decreased $R_m(L)$.  For any given range $L$ there should exist an optimal (key-rate-maximizing) solution for spatial geometry (tiling) of the FBs, power allocation across the FBs, and beam widths. For shorter range $L$ the optimal solution should involve a greater number of FBs, and the number of beams employed should be approximately proportional to $D_{\rm f}$. 

Here, instead of evaluating the optimal rate-maximizing solution as explained above (which is extremely difficult), we find a numerical solution to a constrained optimization problem assuming a square-grid tiling of the FBs in the receiver aperture and restricting our attention to the discrete-variable (DV) laser-light decoy-state BB84 protocol~\cite{lo05decoyqkd}. The rationale behind this is to obtain an {\em achievable} rate-distance envelope for the OGBA transmitter to compare with the ultimate key capacity attainable by employing infinitely many LG (or HG) modes. Since we restrict our attention to DV QKD, we assume that the OGBA transmitter is paired with a single-photon detector (SPD) array at the receiver with square-shaped pixels and unity fill factor with each FB being focused at the center of a detector pixel and there are as many detector pixels as the number of FBs (the optimal number of which is a function of $L$ as discussed above). 

Azimuthal LG modes retain their orthogonality when passed through hard-pupil circular apertures. Thus, generating and separating these modes without any power leaking between them is possible in theory, and has been the subject of much experimental work \cite{mirhosseini13oammodesseparation, lavery12sortingoam}.  The current state of the art is the separation of 25 OAM modes with average efficiency of $>92\%$, as was demonstrated in \cite{mirhosseini13oammodesseparation}.  We compare the QKD rate achievable with our OGBA proposal to what is achievable using ideal separation of azimuthal LG modes as well as the best currently possible.  In the latter case,  we obtained the data for the cross-talk (overlap) between the separated modes (see \cite[Table 4a]{mirhosseini13oammodesseparation}) from the authors of \cite{mirhosseini13oammodesseparation}.  We evaluate performance  assuming ideal photodetectors and no atmospheric extinction.  We find that the achievable rate using our OGBA architecture is at worst $4.4$ dB less than the state-of-the-art azimuthal LG mode separation in \cite{mirhosseini13oammodesseparation} and at worst $8.3$ dB less than the theoretical maximum for entire azimuthal LG mode set, while using hard-pupil transmitter and receiver apertures of same areas and the same center wavelength.  The maximum rate gap occurs because the square-grid OGBA architecture does allow the use of two and three beams; with two square pixels placed side-by-side at the receiver, the gap between the systems employing the state-of-the-art and ideal azimuthal LG mode separation reduces to $2.6$ dB and $6.3$ dB, respectively.  
Current technology for optical communication using orthogonal modes use bulky and expensive components \cite{willner15oam}. While advances in enabling technology could reduce the device size, weight and cost of orthogonal mode generation and separation, our results show that using OAM modes for QKD may not be worth the trouble: the gain in QKD key rate in the near field is modest compared to what can already be obtained by our fairly simple-to-implement OGBA architecture.

This paper is organized as follows: in the next section we introduce the basic mathematics of laser light propagation in vacuum using soft-pupil (Gaussian attenuation) apertures. In Section \ref{sec:lgmodes} we consider the propagation of LG modes using hard-pupil circular apertures, while in Section \ref{sec:gaussian} we discuss the mathematical model of the OGBA architecture that we propose in this paper.  Using the expressions derived in Sections \ref{sec:lgmodes} and \ref{sec:gaussian}, we numerically evaluate the QKD rate using various beam and aperture geometries, and report the results in Section \ref{sec:results}.  We conclude with a discussion of the implications of our results as well as future work, in Section \ref{sec:discussion}.

\section{Bosonic Mode Sets and the Degrees of Freedom of the Photon}
\label{sec:bosonic}

Consider propagation of linearly-polarized, quasimonochromatic light with 
  center wavelength $\lambda$ (that is, a narrow transmission band 
  $\Delta \lambda \ll \lambda$ around the center wavelength) from Alice's 
  transmitter pupil in the $z=0$ transverse plane with a complex-field-unit
  pupil function $A_{\rm T}({\bm \rho})$, ${\bm \rho} \equiv (x,y)$, through 
  a $L$-meter line-of-sight free-space channel, and received by Bob's receiver 
  pupil in the $z=L$ plane with aperture function 
  $A_{\rm R}({\bm \rho^\prime})$, 
  ${\bm \rho^\prime} \equiv (x^\prime,y^\prime)$.
Alice's transmitted field's complex envelope $E_0({\bm \rho},t)$ is multiplied 
  (truncated) by the complex-valued transmit-aperture function 
  $A_{\rm T}({\bm \rho})$, undergoes free-space diffraction over the $L$-meter 
  path, and is truncated by Bob's receiver-aperture function 
  $A_{\rm R}({\bm \rho^\prime})$, to yield the received field 
  $E_L({\bm \rho^\prime},t)$. 
The overall input-output relationship is described by the following 
  linear-system equation:
  \begin{align}
  E_L({\bm \rho^\prime},t)&= \int E_0({\bm \rho},t - L/c) \, h({\bm \rho^\prime}, {\bm \rho}, t) \, \mathrm{d}^2 {\bm \rho},
  \label{eq:fresnel}
  \end{align}
  where the channel's Green's function $h({\bm \rho^\prime}, {\bm \rho}, t)$ is 
  a spatial impulse response. 
We assume vacuum propagation and drop the time argument $t$ from the Green's 
  function:
\begin{align}
\label{eq:vacpropkernel} h({\bm \rho^\prime}, {\bm \rho})& = A_{\rm R}({\bm \rho^\prime}) \, \frac{\exp\left[ik \left(L + |{\bm \rho^\prime}-{\bm \rho}|^2/2L\right)\right]}{i \lambda L} \, A_{\rm T}({\bm \rho}),
\end{align}
where $k = 2\pi/\lambda$.
Normal-mode decomposition of the vacuum-propagation Green's function yields 
  an infinite set of orthogonal input-output spatial-mode pairs (a mode being 
  a normalized spatio-temporal field function of a given polarization), that is,
  an infinite set of non-interfering parallel spatial channels. 
In other words,
\begin{align}
\label{eq:eigenmodes}\int h({\bm \rho^\prime},{\bm \rho}) \Phi_m({\bm \rho})\mathrm{d}^2{\bm \rho} &= \sqrt{\eta_m} \,\phi_m({\bm \rho^\prime}), \, {\rm for}\, m=1,2,\ldots,
\end{align}
  where $\left\{\Phi_m({\bm \rho})\right\}$ forms a complete orthonormal (CON) 
  spatial basis in the transmit-aperture plane before the aperture mask 
  $A_{\rm T}({\bm \rho})$, and $\left\{\phi_m({\bm \rho^\prime})\right\}$ forms
  a CON spatial basis in the receiver-aperture plane after the aperture mask 
  $A_{\rm R}({\bm \rho^\prime})$. 
That is,
\begin{align}
\int\Phi_m({\bm \rho})\Phi_n({\bm \rho})\mathrm{d}^2{\bm \rho}=\delta_{m,n},~\int|\Phi_m({\bm \rho})|^2\mathrm{d}^2{\bm \rho}=1\\
\int\phi_m({\bm \rho})\phi_n({\bm \rho})\mathrm{d}^2{\bm \rho}=\delta_{m,n},~\int|\phi_m({\bm \rho})|^2\mathrm{d}^2{\bm \rho}=1,
\end{align}
where 
  $\delta_{m,n}=\left\{\begin{array}{lr}1&\text{if~}m=n\\0&\text{if~}m\neq n\end{array}\right.$
  is the Kronecker delta function.
Therefore, the singular-value decomposition (SVD) of 
  $h({\bm \rho^\prime},{\bm \rho})$ yields:
\begin{align}
h({\bm \rho^\prime},{\bm \rho})&=\sum_{m=1}^\infty \sqrt{\eta_m} \,\phi_m({\bm \rho^\prime})\Phi_m^\ast({\bm \rho}).
\end{align}
Physically this implies that if Alice excites the spatial mode 
  $\Phi_m({\bm \rho})$, it in turn excites the corresponding spatial mode 
  $\phi_m({\bm \rho^\prime})$ (and no other) within Bob's receiver. 
This specific set of transmitter-plane receiver-plane spatial-mode pairs that 
  form a set of non-interfering parallel channels are the eigenmodes for 
  the channel geometry. 
The fraction of power Alice puts in the mode $\Phi_m({\bm \rho})$ that appears 
  in Bob's spatial mode $\phi_m({\bm \rho^\prime})$ is the modal transmissivity,
  $\eta_m$. 
We assume that the modes are ordered such that
\begin{align}
  1 \ge \eta_1 \ge \eta_2 \ge \ldots \eta_m \ge \ldots \ge 0.
\end{align}
If Alice excites the mode $\Phi_m({\bm \rho})$ in a coherent-state 
  $|\beta\rangle$---the quantum description of an ideal laser-light pulse of 
  intensity $|\beta|^2$ (photons) and phase ${\rm Arg}(\beta)$, then 
  the resulting state of Bob's mode $\phi_m({\bm \rho^\prime})$ is 
  an attenuated coherent state $|\sqrt{\eta_m}\beta\rangle$.
The power transmissivities $\eta_m(\omega)$ are strictly increasing functions 
  of the transmission frequency $\omega = 2\pi c/\lambda$, each increasing from
  $\eta_m = 0$ at $\omega = 0$, to $\eta_m = 1$ at $\omega = \infty$. 

Let us consider Gaussian-attenuation (soft-pupil) apertures with 
\begin{align}
\label{eq:tx_gauss_ap}  A_{\rm T}({\bm \rho}) &= \exp\left[-|{\bm \rho}|^2/r_{\rm t}^2\right] \text{~and}\\ 
\label{eq:rx_gauss_ap}  A_{\rm R}({\bm \rho^\prime}) &= \exp\left[-|{\bm \rho^\prime}|^2/r_{\rm r}^2\right].
\end{align}
For this choice of pupil functions, there are two unitarily-equivalent sets of eigenmodes: the 
   aforementioned Laguerre-Gauss (LG) modes, which have circular symmetry in 
   the transverse plane and are known to carry orbital angular momentum (OAM), 
   and the Hermite-Gauss (HG) modes, which have rectangular symmetry in 
   the transverse plane and do not carry OAM. 
The input LG modes, labeled by the radial index $p=0,1,2,\ldots $ and 
  the azimuthal index $l=0,\pm 1,\pm 2, \ldots$, are expressed using the polar 
  coordinates ${\bm \rho}\equiv(r,\theta)$ as follows:
\begin{align}
\label{eq:inputLG} \Phi_{p,l}(r,\theta)&=\sqrt{\frac{p!}{\pi(|l|+p)!}}\frac{1}{a}\left[\frac{r}{a}\right]^{|l|}\mathcal{L}_p^{|l|}\left(\frac{r^2}{a^2}\right)\exp\left(-\left[\frac{1}{2a^2}+\frac{ik}{2L}\right]r^2+il\theta\right),
\end{align}
where $\mathcal{L}_p^{|l|}(\cdot)$ denotes the generalized Laguerre polynomial 
  indexed  by $p$ and $|l|$.
For completeness of exposition, the input HG modes, labeled by the horizontal 
  and vertical indices $n,m=0,1,2,\ldots$, are expressed using the Cartesian
  coordinates ${\bm \rho}\equiv(x,y)$ as follows:
\begin{align}
\Phi_{n,m}(x,y)&=\frac{1}{a\sqrt{\pi n! m! 2^{n+m}}}H_n\left(\frac{x}{a}\right)H_m\left(\frac{y}{a}\right)\exp\left(-\left[\frac{1}{2a^2}+\frac{ik}{2L}\right][x^2+y^2]\right)
\end{align}
where $H_p(\cdot)$ is the $p^{\text{th}}$ Hermite polynomial.
In the expressions for both LG and HG modes, $a$ is a beam width parameter 
  given by
\begin{align}
\label{eq:a_gauss}a&=\frac{r_{\rm t}}{\sqrt{2}(1+4D_{\rm f})^{1/4}},
\end{align}
where
\begin{align}
D_{\rm f}&=\frac{kr_{\rm t}^2}{4L}\frac{kr_{\rm r}^2}{4L}
\end{align}
is the product of the transmitter-pupil and receiver-pupil Fresnel number 
  products for this soft-pupil vacuum propagation configuration.
Alternatively, $D_{\rm f}={A_\mathrm{t} A_\mathrm{r}}/{(\lambda L)^2}$ when 
  expressed using the transmitter and receiver pupils' areas 
  $A_{\rm t} \equiv \int |A_{\rm T}({\bm \rho})|^2 \mathrm{d}^2{\bm \rho} =  
  \frac{\pi r_{\rm t}^2}{2}$ and 
  $A_{\rm r} \equiv \int |A_{\rm R}({\bm \rho^\prime})|^2 
  \mathrm{d}^2{\bm \rho^\prime} =  \frac{\pi r_{\rm r}^2}{2}$.
The expressions for the output LG and HG modes are given by equations (28) and 
  (24) in \cite{Sha05}, respectively.
The expression for the power-transfer eigenvalues $\eta_q$ for either mode set 
  admits the following simple form:
\begin{align}
\eta_{q}& =\left(\frac{1+2D_\mathrm{f}-\sqrt{1+4D_\mathrm{f}}}{2D_\mathrm{f}}\right)^{q}, \, {\rm for}\, q=1,2,\ldots,
\label{eq:gen_modes}
\end{align}
where $q=2p+|l|+1$ for LG modes, and $q=n+m+1$ for HG modes.
Thus, there are $q$ spatial modes of transmissivity $\eta_q$. 
The LG and HG modes span the same eigenspace, and hence are related by 
  a unitary transformation (a linear mode transformation).

The first mode in both LG or HG mode sets, defined by $p=l=n=m=0$,
  is known as the \emph{Gaussian beam}.
The input Gaussian beam is expressed as follows:
\begin{align}
\label{eq:inputGauss}\Phi_{0,0}(x,y)&=\frac{1}{\sqrt{\pi}a}\exp\left(-\left[\frac{1}{2a^2}+i\frac{k}{2L}\right](x^2+y^2)\right).
\end{align}

\section{LG Modes and Hard-Pupil Circular Apertures}
\label{sec:lgmodes}
Soft-pupil Gaussian apertures used in the preceding section are purely 
  theoretical constructs: while they greatly simplify the mathematics, they are 
  impossible to realize physically.
Let us thus consider hard-pupil circular apertures of areas $A_{\rm t}$ and 
  $A_{\rm r}$, that is,
\begin{align}
\label{eq:ATc} A_{\rm T}({\bm \rho})& = \left\{\begin{array}{ll}1&\text{if~} |{\bm \rho}| \le r_{\rm t}\\0&\text{otherwise}\end{array}\right.,\text{~and}\\
\label{eq:ARc} A_{\rm R}({\bm \rho^\prime})& = \left\{\begin{array}{ll}1&\text{if~} |{\bm \rho^\prime}| \le r_{\rm r}\\0&\text{otherwise}\end{array}\right.
\end{align}
with the corresponding areas defined as
$A_{\rm t} \equiv \int |A_{\rm T}({\bm \rho})|^2 \mathrm{d}^2{\bm \rho} =  \pi r_{\rm t}^2$ and
  $A_{\rm r} \equiv \int |A_{\rm R}({\bm \rho^\prime})|^2 \mathrm{d}^2{\bm \rho^\prime} = \pi r_{\rm r}^2$.
Neither LG nor HG modes form an eigenmode set for these hard-pupil apertures.
Instead, their eigenmodes are prolate spheroidal functions, and 
  the power-transfer eigenvalues $\eta_m(\omega)$, indexed by two integers 
  $m \equiv (m_1, m_2)$, have known, yet quite complicated 
  expressions~\cite{slepian64prolate,slepian65apodization}. 
If the LG (or HG) modes are used as input into the hard-pupil system,
  the output modes are non-orthogonal in general, as the expressions that we
  derive next show.

Employing the vacuum propagation kernel in \eqref{eq:vacpropkernel} with 
  the expression for the input LG mode in \eqref{eq:inputLG},  substituting
  the expressions for the hard circular pupils in \eqref{eq:ATc} 
  and \eqref{eq:ARc}, and re-arranging terms yields:
\begin{align}
\label{eq:phi_out}\phi_{p,l}(r',\theta')&=\frac{\exp[ikL+il\theta']\sqrt{p!}}{i a\lambda L\sqrt{\pi(|l|+p)!}}\int_0^{r_{\rm t}}\int_{0}^{2\pi} \left[\frac{r}{a}\right]^{|l|}\mathcal{L}_p^{|l|}\left[\frac{r^2}{a^2}\right]\exp\left[-\frac{r^2}{2a^2}+\frac{ik}{2L}(r'^2-2rr'\cos\theta)+il\theta\right]r\mathrm{d}\theta\mathrm{d}r,
\end{align}
for $r'\in[0,r_{\rm r}]$ and $\theta'\in[0,2\pi]$, where we first substitute
  $|{\bm \rho^\prime}-{\bm \rho}|^2=r^2+r'^2-2rr'\cos(\theta-\theta')$,
  and then substitute $\theta\rightarrow\theta-\theta'$.
Now, the integral representation of the Bessel function of the first kind
  given in Appendix \ref{app:besselint} allows the following evaluation of
  the integral with respect to $\theta$ in \eqref{eq:phi_out}:
\begin{align}
\label{eq:inner_int}\int_{0}^{2\pi}\exp\left[i\left(l\theta-\frac{krr'\cos\theta}{L}\right)\right]\mathrm{d}\theta=2\pi \exp\left[-\frac{il\pi}{2}\right]J_l\left[\frac{krr'}{L}\right].
\end{align}
Substitution of \eqref{eq:inner_int} into \eqref{eq:phi_out} yields:
\begin{align}
\label{eq:phi_out1}\phi_{p,l}(r',\theta')&=\frac{2\exp\left[ikL+il\theta'+\frac{ikr'^2}{2L}-\frac{il\pi}{2}\right]\sqrt{\pi p!}}{i a\lambda L\sqrt{(|l|+p)!}}\int_0^{r_{\rm t}} \left[\frac{r}{a}\right]^{|l|}\mathcal{L}_p^{|l|}\left[\frac{r^2}{a^2}\right]\exp\left[-\frac{r^2}{2a^2}\right]J_l\left[\frac{krr'}{L}\right]r\mathrm{d}r.
\end{align}
While the Bessel function is not an elementary function, it can be efficiently
  evaluated by a computer (using, e.g., MATLAB).

Now let's evaluate the cross-talk (overlap) between the output modes.
We are interested in the fraction of power transmitted on the mode indexed by 
  $(p,l)$ that is leaked to the mode indexed by $(q,m)$:
\begin{align}
\label{eq:radialcrosstalk}\eta_{\rm L}(p,l,q,m)&=\left|\int_0^{r_{\rm r}}\int_{0}^{2\pi} \phi_{p,l}(r',\theta')\phi_{q,m}^\ast(r',\theta')r'\mathrm{d}\theta'\mathrm{d}r'\right|^2.
\end{align}
Substituting \eqref{eq:phi_out1}, we note that evaluation of the integral
  with respect to $\theta'$ yields: 
  $\int_{0}^{2\pi}\exp\left[i(l-m)\theta'\right]\mathrm{d}\theta'=2\pi\delta_{l,m}$.
Thus, while the radial LG modes are clearly non-orthogonal, the azimuthal LG 
  modes retain their orthogonality when passed through hard-pupil circular 
  apertures.
However, azimuthal LG modes are unlikely to be perfectly separated in 
  the near future.
The current state-of-the-art experiments have been able to achieve 
  $\eta_{\rm L}=7.9\pm0.7\%$ averaged across the 25 modes spanning 
  $l=-12,-11,\ldots,12$ \cite{mirhosseini13oammodesseparation}; we evaluate
  the QKD rate for such a system using the cross-talk data from 
  these experiments.

\section{Gaussian Beam Array and Hard-Pupil Square Apertures}
\label{sec:gaussian}
Our OGBA architecture employs a square transmitter aperture. 
The receiver aperture is composed from square pixels of equal size.
Gaussian beams are directed from the transmitter to the square pixels using
  linear phase tilts as in \eqref{eq:phasetilt}.
The hard square pupils of areas $A_{\rm t}$ and $A_{\rm r}$ are given by:
\begin{align}
        \label{eq:ATs} A_{\rm T}({\bm \rho})& = \left\{\begin{array}{ll}1&\text{if~} |x|,|y| \le l_{\rm t}/2\\0&\text{otherwise}\end{array}\right.,\text{~and}\\
\label{eq:ARs} A_{\rm R}({\bm \rho^\prime})& = \left\{\begin{array}{ll}1&\text{if~} |x'|,|y'| \le l_{\rm r}/2\\0&\text{otherwise}\end{array}\right..
\end{align}
The corresponding areas are defined as
$A_{\rm t} \equiv \int |A_{\rm T}({\bm \rho})|^2 \mathrm{d}^2{\bm \rho} =  l_{\rm t}^2$ and
  $A_{\rm r} \equiv \int |A_{\rm R}({\bm \rho^\prime})|^2 \mathrm{d}^2{\bm \rho^\prime} = l_{\rm r}^2$.

For simplicity of exposition, we ignore the linear phase tilt of the input 
  Gaussian beam (and the corresponding offset of the output Gaussian beam),
  and derive the expression for the beam centered on the central pixel of 
  the output aperture (in fact, while the implementation of the Gaussian beam 
  array would use the linear phase tilts, we do not need to explicitly consider 
  them in the analysis that follows). 
The beams directed at each pixel have intensity $|\alpha|^2$, which we optimize
  in the next section.
Employing the vacuum propagation kernel in \eqref{eq:vacpropkernel} with 
  the expression for the input Gaussian beam $\Phi_{0,0}(x,y)$ in 
  \eqref{eq:inputGauss}, substituting expressions for the hard square pupils in 
  \eqref{eq:ATs} and \eqref{eq:ARs}, and re-arranging terms yields the 
  following:
\begin{align}
\phi_{0,0}(x',y')&=\frac{\sqrt{\pi}a\exp\left[ikL-(x'^2+y'^2)\left(\frac{a^2k^2}{2L^2}-\frac{ik}{2L}\right)\right]}{2i\lambda L}\left(\erf\left[\frac{l_{\rm t}}{\sqrt{2}a}-\frac{iakx'}{\sqrt{2}L}\right]+\erf\left[\frac{l_{\rm t}}{\sqrt{2}a}+\frac{iakx'}{\sqrt{2}L}\right]\right)\nonumber\\
&\phantom{=}\times\left(\erf\left[\frac{l_{\rm t}}{\sqrt{2}a}-\frac{iaky'}{\sqrt{2}L}\right]+\erf\left[\frac{l_{\rm t}}{\sqrt{2}a}+\frac{iaky'}{\sqrt{2}L}\right]\right)\\
\label{eq:phiGauss}&=\frac{2\sqrt{\pi}a\exp\left[ikL-(x'^2+y'^2)\left(\frac{a^2k^2}{2L^2}-\frac{ik}{2L}\right)\right]}{i\lambda L}\mathfrak{Re}\left[\erf\left[\frac{l_{\rm t}}{\sqrt{2}a}+\frac{iakx'}{\sqrt{2}L}\right]\right]\mathfrak{Re}\left[\erf\left[\frac{l_{\rm t}}{\sqrt{2}a}+\frac{iaky'}{\sqrt{2}L}\right]\right],
\end{align}
where $\erf(x)=\frac{2}{\sqrt{\pi}}\int_0^xe^{-t^2}\mathrm{d}t$ is 
  the error function and the simplification in \eqref{eq:phiGauss} is because
  of the symmetry of $\erf(\cdot)$ as explained in Appendix \ref{app:erf}.
While the error function is not an elementary function, it can be efficiently
  evaluated by a computer (using, e.g., the Faddeeva Package \cite{faddeeva} 
  which includes a wrapper for MATLAB).

\begin{figure}[t]
\centering
\subfloat[Centered single pixel]{\includegraphics[width=0.29\columnwidth]{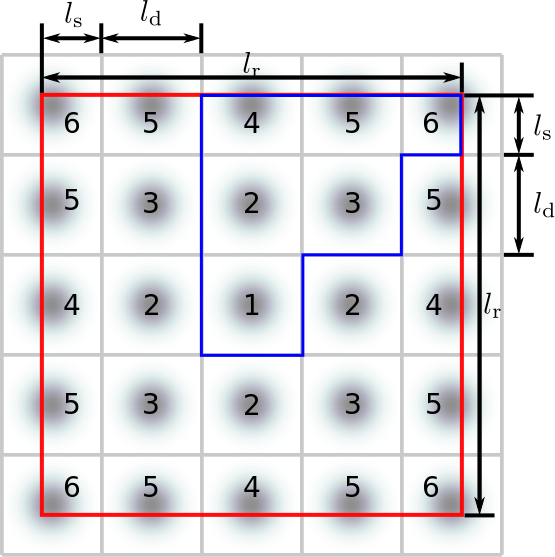}\label{fig:config_1}}
\hfill
\subfloat[Centered $2\times2$ pixel cluster]{\includegraphics[width=0.29\columnwidth]{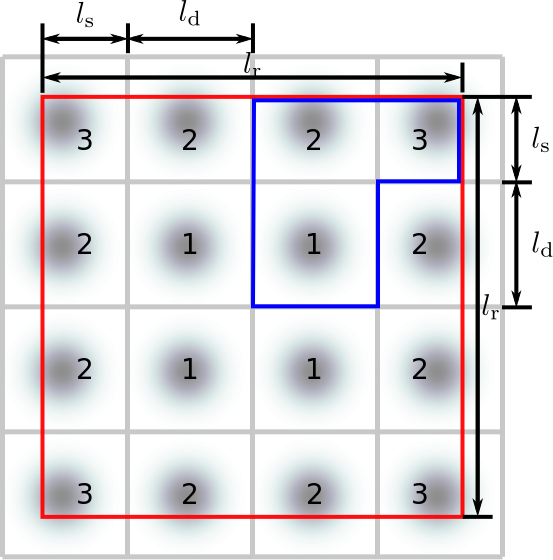}\label{fig:config_2}}
\subfloat[$1\times2$ pixel array]{\includegraphics[width=0.41\columnwidth]{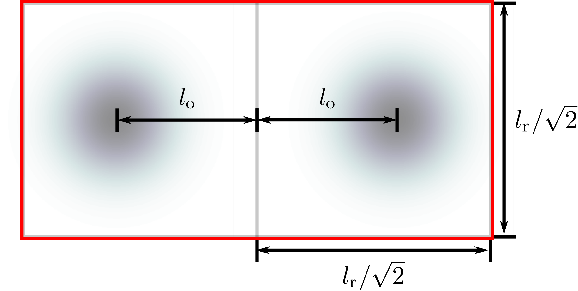}\label{fig:config_3}}
\caption{The illustration of configurations for the layout of receiver pixels. 
  The aperture border is marked red while pixel
  borders are marked gray.  
  In \protect\subref{fig:config_1} and \protect\subref{fig:config_2} the pixels
  on the edge of the apertures are cut off
  to fit, even though the beams are directed at their centers as shown.
  Because of circular symmetry of Gaussian beams and their equal intensity, 
  identically numbered pixels 
  experience identical interference from other beams, which makes them 
  equivalent in our model.
  We thus limit out calculations to the set of pixels outlined in blue.
  In \protect\subref{fig:config_3}, the beams are horizontally offset from 
  the center.}
\label{fig:config}
\end{figure}

Suppose that the receiver aperture is constructed using square 
  $l_{\rm d}\times l_{\rm d}$ m pixels.
We consider two configurations for the layout of these pixels on the square 
  aperture as illustrated in Figure \ref{fig:config}:
\begin{enumerate}
\item \label{item:center} a pixel in the center of the aperture as shown in 
  Figure \ref{fig:config_1}; 
\item \label{item:cluster} a $2\times2$ pixel cluster in the center of 
  the aperture as shown in Figure \ref{fig:config_2}; and
\item a $1\times 2$ pixel array with two square 
  pixels placed side-by-side as shown in Figure \ref{fig:config_3}.
\end{enumerate}
Consider configurations \ref{item:center} and \ref{item:cluster}.
We optimize the length of the pixel $l_{\rm d}$ when computing the QKD rate.
Unless $l_{\rm r}/l_{\rm d}$ is an integer, the pixels at the edges of 
  the aperture are cut off to fit into the aperture.
While these pixels are either $l_{\rm d}\times l_{\rm s}$ m rectangles on 
  the edges of the aperture or $l_{\rm s}\times l_{\rm s}$ m squares on 
  the corner, for simplicity we still direct the beams at the centers of 
  the hypothetical full $l_{\rm d}\times l_{\rm d}$ m pixels that are cut off 
  by the edge of the aperture.
The circular symmetry of the Gaussian beam allows us to limit our
  calculations to a set of pixels forming octants illustrated in
  Figure \ref{fig:config}, as interference profiles for the
  corresponding pixels in other octants are identical.
The total QKD rate is computed by summing the products of the contribution from 
  each of these pixels with the total number of identical pixels.

Using paraxial approximation, the fraction of power captured by a full
  (interior) $l_{\rm d}\times l_{\rm d}$ m pixel from a Gaussian beam 
  focused on its center is:
\begin{align}
\label{eq:eta}\eta&=\int_{-l_{\rm d}/2}^{l_{\rm d}/2}\int_{-l_{\rm d}/2}^{l_{\rm d}/2}|\phi_{0,0}(x',y')|^2\mathrm{d}x'\mathrm{d}y'.
\end{align}
The fraction of power captured by a partial (edge) pixel is obtained by
  appropriately adjusting the limits of integration in \eqref{eq:eta}.
Since Gaussian beam is circularly symmetric, the cross-talk from another beam
  that is focused on a pixel whose center is located $n$ pixels either to 
  the left or to the right and $m$ pixels either above or below is expressed 
  similarly:
\begin{align}
\label{eq:crosstalkGauss}\eta_{\rm L}(n,m)&=\int_{l_{\rm d}\left(m-\frac{1}{2}\right)}^{l_{\rm d}\left(m+\frac{1}{2}\right)}\int_{l_{\rm d}\left(n-\frac{1}{2}\right)}^{l_{\rm d}\left(n+\frac{1}{2}\right)}|\phi_{0,0}(x',y')|^2\mathrm{d}x'\mathrm{d}y'.
\end{align}
Again, the cross-talk from another beam captured by a partial (edge) pixel is 
  obtained by appropriately adjusting the limits of integration in 
  \eqref{eq:crosstalkGauss}.
The total contribution of interference from cross-talk to noise afflicting 
  the detector at the pixel that is $u$ pixels to the right and $v$ pixels above
  the bottom-left pixel is calculated by summing each interfering beam's 
  cross-talk given in \eqref{eq:crosstalkGauss} and multiplying by the beam
  intensity $|\alpha|^2$:
\begin{align}
P_{\rm L}(u,v)&=\sum_{\myatop{n,m\in\{0,\ldots,\lceil l_{\rm r}/l_{\rm d}\rceil-1\}}{n\neq u\lor m\neq v}} |\alpha|^2\eta_{\rm L}(|u-n|,|v-m|).
\end{align}

In configuration 3 two $l_{\rm s}\times l_{\rm s}$ square pixels are placed 
  side-by-side, where $l_{\rm s}=l_{\rm r}/\sqrt{2}$.
The beams are vertically centered on the corresponding square pixel but
  can be offset horizontally.
When computing QKD rate, we optimize the distance from the center of 
  the aperture for both beams $l_{\rm o}$.
The fraction of power captured by each pixel and the cross-talk can be 
  calculated by appropriately setting the limits of integration in 
  \eqref{eq:eta} and \eqref{eq:crosstalkGauss}.

\section{Results}
\label{sec:results}

\begin{figure}[t]
\centering
\subfloat[QKD rates for all systems]{\includegraphics[width=0.45\columnwidth]{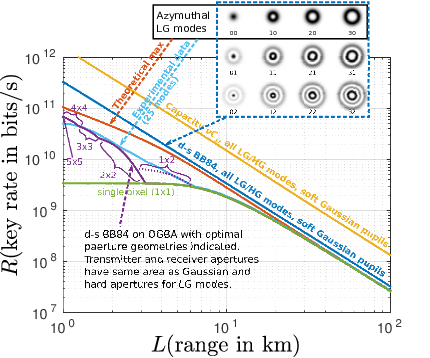}\label{fig:mult_modes}}
\hfill
\subfloat[Comparison of OGBA and azimuthal LG modes]{\includegraphics[width=0.45\columnwidth]{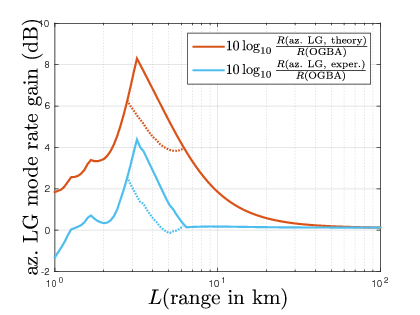}\label{fig:comp_rates}}
\caption{QKD rate for various beam and aperture geometries, and the comparison
  of the QKD rates achieved for hard apertures of equal areas using overlapping
  Gaussian beam array (OGBA, purple plot) and azimuthal LG mode sets (red plot
  for the theoretical capacity of infinitely many azimuthal LG modes, and 
  light blue plot for the maximum rate achievable by the state-of-the-art 
  system in \cite{mirhosseini13oammodesseparation} that separates 25 azimuthal
  LG modes).} 
\end{figure}

We plot our results in Figure \ref{fig:mult_modes}.
We assume vacuum propagation without any extinction losses and turbulence;
  the losses and cross-talk induced by the channel are solely from diffraction.
Our repetition rate is $\nu=10^{10}$ modes/s.
The yellow line is the capacity of QKD system (see discussion of \eqref{eq:Cs} 
  in Appendix \ref{app:decoyBB84}) that employs both polarizations, full set of 
  orthogonal spatial modes and soft Gaussian apertures.
That is, it plots
\begin{align}
\nu C_s=-2\nu\sum_{q=1}^\infty q\log_2(1-\eta_q),
\end{align}
where $\eta_q$ is given by \eqref{eq:gen_modes}.

Next we examine the performance of the decoy state BB84 protocol that is
  reviewed in Appendix \ref{app:decoyBB84}.
All of these results are for apertures with total area 
  $A=0.005\pi~\text{m}^2$, i.e., the  effective area of a soft-pupil Gaussian 
  aperture (as defined in Section \ref{sec:bosonic}) with $r=0.1$ m and the
  hard-pupil circular aperture of radius $r\approx 0.07$ m.
The areas of transmitter and receiver apertures are equal.
Our operating wavelength is $\lambda=1.55~\mu$m.
We assume dark click probability $p_{\rm d}=10^{-6}$, unity detector quantum 
  efficiency $\eta_{\rm d}=1$, visibility $V=0.99$ (i.e., the probability that
  the beam splitter directs the pulse according to the bases chosen by Bob),
  and availability of capacity-achieving channel codes 
  (i.e., error correction code efficiency $f_{\mathrm{leak}}=1$).
We optimize QKD rate $R(L)$ that is calculated in Appendix \ref{app:decoyBB84}
  over the intensity of Alice's pulses $|\alpha|^2$.
The blue curve plots the QKD rate of a system employing the entire orthogonal 
  spatial mode set with soft-pupil Gaussian apertures.
To obtain this rate, we optimize over the intensity of Alice's pulses 
  $|\alpha|^2$.

The soft-pupil Gaussian apertures are mathematically convenient devices,
  however, they are not realizable in practice.
We thus turn our attention to hard-pupil apertures with the same area.
First we examine azimuthal LG modes.
The red curve plots the maximum rate achievable in theory using this mode set
  when hard-pupil circular apertures are used.
There is no cross-talk between the modes since they retain their orthogonality.
We optimize over the beam width $a$ and intensity $|\alpha|^2$ using the entire
  infinite set of LG modes, however noting that modes with high index couple 
  only an insignificant portion of power from the transmitter to the receiver,
  and thus are not used at long distances.
We also evaluate the theoretical performance of the decoy state BB84 QKD 
  protocol using the data from the experimental system for separating
  25 azimuthal LG modes indexed from -12 to 12 
  \cite{mirhosseini13oammodesseparation}, and plot the results with the light
  blue curve.
This is the current state-of-the-art in azimuthal LG mode separation.
Because of various imperfections inherent in physical systems, there is 
  cross-talk between modes in these experiments, as depicted in 
  \cite[Figure 4a]{mirhosseini13oammodesseparation};
  we obtained these data from the authors.
We treat the erroneous counts from cross-talk as we treat the detector dark 
  counts.
The only source of loss is diffraction; we assume that there are no losses 
  incurred in mode separation (even though they may be substantial) as
  well as through extinction and turbulence.
In order to make a fair comparison between various systems,
  we normalize the cross-talk probabilities over the modes which couple 
  significant power to the receiver (i.e., modes that we use).\footnote{For 
  example, suppose that mode separator couples 80\% of the received input power 
  from the mode indexed 0 to mode 0, 7\% to each of the modes indexed -1 
  and +1, and 3\% to each of the modes indexed -2 and +2.
  If we only use modes indexed -1, 0, and 1, then we normalize the
  cross-talk probabilities so that in our calculations the mode separator
  couples 85.1\% of the received input power from the mode indexed 0 to mode
  0 and 7.45\% to each of the modes indexed -1 and 1.}
This normalization, while not ideal, avoids treating photons sent on the zeroth
  mode as lost to cross-talk in separation when only the zeroth mode is
  used (the case when $L$ is large).
We optimize the QKD rate $R(L)$ over the beam width $a$ and intensity 
  $|\alpha|^2$. 
The dip on the left side of the light blue curve (around $L=1$ km) is because
  the experiment was limited to 25 modes, more modes would improve the rate
  in that regime.

\begin{figure}[t]
\centering
\subfloat[Beam width $a$]{\includegraphics[width=0.3\columnwidth]{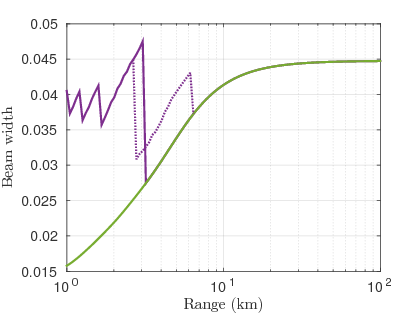}\label{fig:a}}
\hfill
\subfloat[Pixel width $l_{\rm d}$]{\includegraphics[width=0.3\columnwidth]{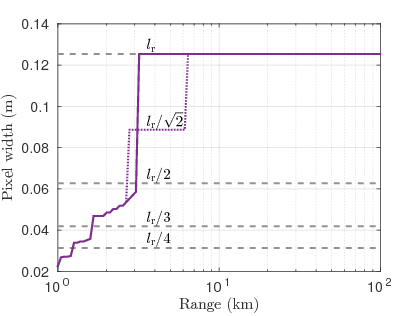}\label{fig:det_length}}
\hfill
\subfloat[Intensity $|\alpha|^2$]{\includegraphics[width=0.3\columnwidth]{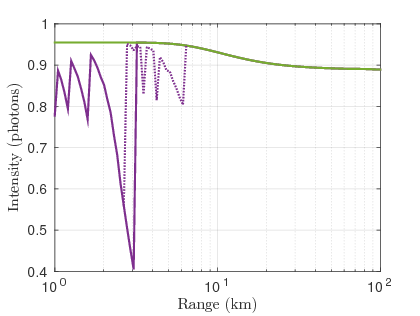}\label{fig:mu}}
\caption{Parameters yielding optimal performance of the QKD system using 
  OGBA (green curves) and a single Gaussian FB (light blue curves).}
\label{fig:params}
\end{figure}

The purple curve in Figure \ref{fig:mult_modes} plots the QKD rate using 
  an optimal number of focused beams with an optimal choice of their overlap 
  at the receiver aperture plane (the optimal overlap is range-dependent).
The transmitter is equipped with an $l_{\rm t}\times l_{\rm t}$ hard-pupil
  square aperture, where the length of the transmitter aperture is equal to the
  total length of the receiver aperture $l_{\rm t}=l_{\rm r}$.
The receiver geometry is as shown in Figures \ref{fig:config_1} and 
  \ref{fig:config_2}.
For each beam we employ the same beam width $a$ and intensity $|\alpha|^2$,
  optimizing over those variables as well as length of the side of the full 
  interior pixel $l_{\rm d}$.
The dashed purple curve plots the maximum QKD rate achievable using  
  $1\times 2$ receiver pixel setup described in Figure
  \ref{fig:config_3} (we keep the $l_{\rm t}\times l_{\rm t}$ hard-pupil
  square transmitter aperture).
Again, we optimize over the beam width $a$ and intensity $|\alpha|^2$,
  however, instead of pixel side length $l_{\rm d}$ (which is set to 
  $l_{\rm r}/\sqrt{2}$), we optimize over the beam offset $l_{\rm o}$.
While practical systems have space between detector pixels, for simplicity we
  assume a unity-fill factor single-photon square detector array
  with each beam focused at the center of one detector pixel (except in the
  $1\times2$ configuration). 
The optimal values of $a$, $l_{\rm d}$, and $|\alpha|^2$ are plotted in Figure
  \ref{fig:params}.
The light blue curve plots QKD rate that employs a single Gaussian FB
  and square apertures (we plot the optimal beam width and intensity
  in Figures \ref{fig:a} and \ref{fig:mu}, respectively).
We provide the comparison of the QKD rates achieved for hard apertures of equal
  areas using OGBA and LG mode sets in Figure \ref{fig:comp_rates}.

\section{Discussion and Future Work}
\label{sec:discussion}

The primary takeaways from these results are: 
\begin{enumerate}
\item One can potentially gain between 1 to 2 orders of magnitude in key rate 
  in the near-field propagation regime (e.g., over a $1$ km link using $7$ cm 
  radii apertures at $1.55~\mu$m center wavelength) by using multiple spatial modes. But,
\item QKD using orthogonal azimuthal LG modes may not be worth it given 
  the hardware complexity associated with generating and separating these 
  spatially overlapping orthogonal modes. 
  In this paper, we proposed an overlapping Gaussian beam array (OGBA) 
  architecture, which uses an array of focussed Gaussian beams with 
  an optimized beam geometry.
  OGBA architecture can yield most of the spatial-multiplexing gain in the
  QKD rate in the near field afforded by the use of azimuthal LG modes. 
  As shown in Figure \ref{fig:comp_rates}, the rate gain from using azimuthal LG
  modes over our OGBA architecture is modest: at most $6.3$ dB in theory if 
  the entire azimuthal LG mode set is used with perfect separation and at most 
  $2.6$ dB with the current state-of-the-art azimuthal LG mode separation 
  implemented in the laboratory (without accounting for any losses introduced 
  by the mode separation process).
  The losses associated with generating and separating these modes will
  likely offset this rate improvement. 
\end{enumerate}
Furthermore, the performance of the OGBA (the green curve)
  might improve further if we use hexagonally-packed beam spots as opposed to 
  using a square grid. 
However, we have not examined that yet. 
Finally, in the near-field regime, CV QKD can improve rate substantially over the DV BB84 protocol since the CV scheme can leverage effectively a high-order constellation in the low-loss regime. Therefore, it would be instructive to evaluate an OGBA architecture employing CV QKD with a heterodyne detection array. 

We assumed vacuum propagation in the results reported in this paper. We are extending them to account for the atmospheric turbulence in the ongoing work. Clearly, turbulence will adversely affect all systems. It is known to break the orthogonality of the azimuthal LG modes \cite{chandrasekaran14pieturb1}. While the classical and private capacities of systems using multiple HG, LG, and FB modes are similar in turbulence \cite{chandrasekaran14pieturb2}, the effect of turbulence on the QKD systems using (or not using) adaptive optics at the transmitter and/or the receiver is still unclear.

\begin{acknowledgments}
The authors are grateful to Mohammad Mirhosseini, Mehul Malik, Zhimin Shi, and 
  Robert Boyd for graciously providing the data plotted in
  \cite[Figure 4]{mirhosseini13oammodesseparation}, as well as answering 
  question about their experiment.
\end{acknowledgments}

\appendix

\section{Useful Integral Representation of the Bessel Function of the first kind $J_n(z)$}
\label{app:besselint}
Eq.~(8.411.1) in \cite{gr07tables} gives the following integral representation 
  of Bessel function of the first kind:
\begin{align}
\label{eq:intJ}J_n(z)&=\frac{1}{2\pi}\int_{-\pi}^{\pi}e^{-in\theta+iz\sin\theta}\mathrm{d}\theta,
\end{align}
where $n$ is an integer.
We perform several substitutions to obtain the form of this integral that is
  useful to us.
First, substitute $\theta\rightarrow-\theta$:
\begin{align}
J_n(z)&=\frac{1}{2\pi}\int_{-\pi}^{\pi}e^{in\theta-iz\sin\theta}\mathrm{d}\theta.
\end{align}
Now substitute $\theta\rightarrow\theta+\frac{\pi}{2}$ and split the resulting
  integral:
\begin{align}
J_n(z)&=\frac{e^{in\pi/2}}{2\pi}\int_{-3\pi/2}^{\pi/2}e^{in\theta-iz\cos\theta}\mathrm{d}\theta\\
\label{eq:splitint}&=\frac{e^{in\pi/2}}{2\pi}\int_{-3\pi/2}^{0}e^{in\theta-iz\cos\theta}\mathrm{d}\theta+\frac{e^{in\pi/2}}{2\pi}\int_{0}^{\pi/2}e^{in\theta-iz\cos\theta}\mathrm{d}\theta.
\end{align}
Now, since $e^{in(\theta-2\pi)}=e^{in\theta}$ for integer $n$, and 
  $\cos(\theta-2\pi)=\cos(\theta)$, substitution $\theta\rightarrow\theta-2\pi$
  into the first integral in \eqref{eq:splitint} only changes its limits,
  yielding the form we need:
\begin{align}
J_n(z)&=\frac{e^{in\pi/2}}{2\pi}\int_{\pi/2}^{2\pi}e^{in\theta-iz\cos\theta}\mathrm{d}\theta+\frac{e^{in\pi/2}}{2\pi}\int_{0}^{\pi/2}e^{in\theta-iz\cos\theta}\mathrm{d}\theta\\
&=\frac{e^{in\pi/2}}{2\pi}\int_{0}^{2\pi}e^{in\theta-iz\cos\theta}\mathrm{d}\theta.
\end{align}

\section{Useful Simplification Involving the Symmetry of Error Function}
\label{app:erf}
Let $g(u,v)=\erf(u+iv)+\erf(u-iv)$.
Now,
\begin{align}
g(u,v)&=\mathfrak{Re}[\erf(u+iv)]+i\mathfrak{Im}[\erf(u+iv)]+\mathfrak{Re}[\erf(u-iv)]+i\mathfrak{Im}[\erf(u-iv)]\\
\label{eq:symerf}&=\mathfrak{Re}[\erf(u+iv)]+i\mathfrak{Im}[\erf(u+iv)]+\mathfrak{Re}[\erf^\ast(u+iv)]+i\mathfrak{Im}[\erf^\ast(u+iv)]\\
\label{eq:conj}&=2\mathfrak{Re}[\erf(u+iv)],
\end{align}
where in \eqref{eq:symerf} we use the fact that $\erf(x^\ast)=\erf^\ast(x)$
  and \eqref{eq:conj} follows from the definition of complex conjugation.

\section{Review of Decoy State Quantum Key Distribution}
\label{app:decoyBB84}
Here we review the decoy state discrete variable BB84 QKD protocol 
  \cite{lo05decoyqkd}, borrowing the development of the key generation rate
  expression from \cite[Section IV.B.3]{scarani09rmpQKD}.
Suppose that Alice transmits pulses to Bob at the rate of $\nu$ Hz.
The lower bound for the rate of secure key generation from these pulses is:
\begin{align}
\label{eq:initQKDrate}R&=I_{\rm AB}-\min(I_{\rm AE},I_{\rm BE})\text{~bits/mode},
\end{align}
where $I_{\rm AB}$ denotes the information shared between Alice and Bob,
  while $I_{\rm AE}$ and $I_{\rm BE}$ denote the information captured  
  by eavesdropper Eve from Alice and Bob, respectively.
Privacy amplification aims to destroy Eve's information, sacrificing part of
  the information in the process (hence subtraction in \eqref{eq:initQKDrate}).
We take the minimum of $I_{\rm AE}$ and $I_{\rm BE})$ in \eqref{eq:initQKDrate}
  since Alice and Bob choose the reference set of pulses on which Eve has least 
  information.
The QKD rate in bits/second is then $\nu R$.
For lossy bosonic channels, $R\leq C_s$ \cite{pirandola15QKDcap}, with QKD 
  capacity given by:
\begin{align}
\label{eq:Cs}C_s&=-\log_2(1-\eta) \text{~bits/mode},
\end{align}
  where $\eta$ captures all losses, which include the diffraction described in
  the previous sections, as well as atmospheric losses and detector 
  inefficiency.

Alice transmits a sequence of polarized laser pulses with 
  average intensity $|\alpha|^2$ photons per pulse.
Following the standard BB84 protocol, polarization is chosen by first randomly
  selecting one of two non-orthogonal polarization bases (rectilinear or 
  diagonal), and then encoding a random bit in the selected bases.
Bob randomly chooses one of two polarization bases in which to measure
  the received pulse.
When Alice and Bob select the same bases, Alice's pulse
  is directed to one of two detectors via a polarizing beam splitter and
  ideally only the detector corresponding to the transmitted bit can click,
  registering the detection event (we discuss the non-ideal case later).
When the bases are not the same, either detector can click with equal 
  probability.
We call Bob's detector ``correct'' when it corresponds to Alice's basis choice,
  otherwise we call the detector ``incorrect.''
The probability of a click from a signal pulse when the bases match is:
\begin{align}
p_{\rm p}&=1-e^{-\eta|\alpha|^2}.
\end{align}
In the decoy state BB84 protocol, Alice changes the value of the intensity
  $|\alpha|^2$ randomly from one pulse to the other; she reveals the list of
  values she used at the end of the exchange of transmissions.
This prevents Eve from adapting her attack to Alice's state, and allows Alice 
  and Bob to estimate their parameters in post-processing.

The probability of a click in one of the detectors from either the received 
  pulse or a dark click is:
\begin{align}
p_{\rm r}&=p_{\rm p}(1-p_{\rm d})+2(1-p_{\rm p})p_{\rm d}(1-p_{\rm d}),
\end{align}
where $p_{\rm d}$ is the probability of a dark click.
When pulse is not detected, an error can occur only because of a dark click 
  in the incorrect detector.
The probability of this event is $p_{\rm d}(1-p_{\rm d})(1-p_{\rm p})$.
When the pulse is received, non-idealities of the polarizing beam splitter
  can result in a click in the erroneous detector.
These non-idealities are captured by the visibility parameter $V$, which 
  is effectively the probability that the beam splitter directs the pulse 
  according to the bases chosen by Bob.
Since an incorrect bases choice results in a click happening with equal 
  probability in one of the detectors, the probability of an erroneous click
  with pulse received is $\frac{1}{2}(1-V)p_{\rm p}(1-p_{\rm d})$.
Combining the above probabilities, the quantum bit error rate is:
\begin{align}
Q&=\frac{\frac{1}{2}(1-V)p_{\rm p}(1-p_{\rm d})+p_{\rm d}(1-p_{\rm d})(1-p_{\rm p})}{p_{\rm r}}.
\end{align}
The rate at which Bob can extract information from the clicks at his
  detectors is thus:
\begin{align}
I_{\rm AB}&=1-f_{\rm leak}h_2(Q),
\end{align}
where $h_2(Q)=-Q\log_2Q-(1-Q)\log_2(1-Q)$ is the binary entropy function, 
  $1-h_2(Q)$ is the expression for the Shannon capacity of the binary symmetric 
  channel, and $f_{\rm leak}$ is the efficiency of the error correction code
  (ECC) used by Alice and Bob.

Now let's study the amount of information about the key collected by Eve 
  $I_{\rm E}=\min(I_{\rm AE},I_{\rm BE})$.
She only gains information when photons are transmitted, and provided that Bob 
  detects
  the photon that she forwarded (thus, when Alice does not send a photon but Bob
  detects a dark click, Eve does not obtain any information about the key).
If Alice sends a single photon pulse, Eve has to introduce an error if she is
  to obtain any information.
In this case Eve gains $h_2(\epsilon_1)$ bits of information, where 
  $\epsilon_1$ is the probability of error event when Alice transmits
  a single photon.
Alice transmits a single photon with probability $|\alpha|^2e^{-|\alpha|^2}$, 
  and a detection event occurs at one of the detectors with 
  probability 
\begin{align}
p_{\rm r_1}=|\alpha|^2e^{-|\alpha|^2}(\eta+2(1-\eta)p_{\rm d})(1-p_{\rm d}).
\end{align}
Conditioned on the event that a click occurs in one of Bob's detectors,
  the probability becomes:
\begin{align}
\label{eq:y_1}y_1=\frac{p_{\rm r_1}}{p_{\rm r}}=\frac{|\alpha|^2e^{-|\alpha|^2}(\eta+2(1-\eta)p_{\rm d})}{p_{\rm p}+2(1-p_{\rm p})p_{\rm d}}.
\end{align}
The probability of Alice transmitting one photon and a click occurring 
  in the incorrect detector is:
\begin{align}
p_{\rm r_1^w}&=|\alpha|^2e^{-|\alpha|^2}(1-\eta)p_{\rm d}(1-p_{\rm d}).
\end{align}
Conditioning on the event that Alice transmits a single photon and
  a detection event occurs at one of the detectors yields:
\begin{align}
\epsilon_1&=\frac{p_{\rm r_1^w}}{p_{\rm r_1}}=\frac{(1-\eta)p_{\rm d}}{\eta+2(1-\eta)p_{\rm d}}.
\end{align}
For multi-photon pulses, photon number splitting is an optimal attack, in which
  Eve forwards one photon to Bob and keeps the others.
She gains one bit from the photons she keeps when there is a click in one
  of Bob's detectors.
The probability of a click in one of the detectors when Alice transmits more 
  than one photons is $1-y_0-y_1$ where $y_1$ is given by \eqref{eq:y_1} and
  $y_0$ is the probability of a click in one of the detectors when Alice does
  not transmit a photon given that a click occurred.
Since Alice sends no photons with probability $e^{-|\alpha|^2}$,
  the probability of a click in one of the detectors when Alice does
    not transmit a photon is:
\begin{align}
p_{\rm r_0}&=2p_{\rm d}(1-p_{\rm d})e^{-|\alpha|^2}.
\end{align}
Conditioning on the event that a click occurs in one of Bob's detectors,
  we obtain:
\begin{align}
y_0&=\frac{p_{\rm r_0}}{p_{\rm r}}=\frac{2p_{\rm d}e^{-|\alpha|^2}}{p_{\rm p}(1-p_{\rm d})+2(1-p_{\rm p})p_{\rm d}}
\end{align}
Therefore,
\begin{align}
I_{\rm E}&=y_1h_2(\epsilon_1)+(1-y_0-y_1)\\
&=1-(y_0+y_1(1-h_2(\epsilon_1))).
\end{align}

The expression for the QKD rate is thus:
\begin{align}
R&=\max[0,p_{\rm r}((1-f_{\rm leak}h_2(Q))-(1-(y_0+y_1(1-h_2(\epsilon_1)))))]\\
\label{eq:R}&=\max[0,p_{\rm r}(y_0+y_1(1-h_2(\epsilon_1))-f_{\rm leak}h_2(Q))] \text{~bits/mode}.
\end{align}
We note that in the numerical optimization performed in Section 
  \ref{sec:results} we use a version of \eqref{eq:R} without taking the 
  maximum.
Allowing negative rate allows MATLAB's \texttt{fmincon} function to construct
  the gradient over the entire space of optimization variables.

\end{document}